\documentclass[
 reprint,
 amsmath,amssymb,
 aps,
floatfix,
]{revtex4-1}

\usepackage{graphicx}
\usepackage{dcolumn}
\usepackage{bm}

\begin{document}
 
\preprint{APS/123-QED}

\title
{Carbon nanotube array as a van der Waals two-dimensional hyperbolic material}
\author{R.G. Polozkov}
\email{polozkov@itmo.ru}%
\affiliation{
ITMO University, Saint Petersburg, 197101, Russia
}%

\author{N.Y. Senkevich}%
\affiliation{
ITMO University, Saint Petersburg, 197101, Russia
}%


\author{S. Morina}%
\affiliation{
Science Institute, University of Iceland, Dunhagi 3, IS-107, Reykjavik, Iceland
}%

\author{P. Kuzhir}%
\affiliation{
Belarusian State University, Institute for Nuclear Problems, Bobruiskaya 11, Minsk 220030, Belarus
}%
\affiliation{
Tomsk State University, 36 Lenin Ave, Tomsk 634050, Russia
}%

\author{M.E. Portnoi}%
\affiliation{
ITMO University, Saint Petersburg, 197101, Russia
}
\affiliation{
School of Physics, University of Exeter, Stocker Road, Exeter EX4 4QL, United Kingdom
}
\affiliation{International Institute of Physics, Universidade Federal do Rio Grande do Norte, Natal-RN 59078-970, Brazil
}%

\author{I.A. Shelykh}%
\affiliation{
ITMO University, Saint Petersburg, 197101, Russia
}
\affiliation{
Science Institute, University of Iceland, Dunhagi 3, IS-107, Reykjavik, Iceland
}%

\date{\today}

\begin{abstract}
We use an ab-initio approach to design and study a novel two-dimensional material - a planar array of carbon nanotubes separated by an optimal distance defined by the van der Waals interaction. We show that the energy spectrum for an array of quasi-metallic nanotubes is described by a strongly anisotropic hyperbolic dispersion and formulate a model low-energy Hamiltonian for its semi-analytical treatment. Periodic-potential-induced lifting of the valley degeneracy for an array of zigzag narrow-gap nanotubes leads to the band gap collapse. In contrast, the band gap is opened in an array of gapless armchair tubes. These unusual spectra, marked by pronounced van Hove singularities in the low-energy density of states, open the opportunity for interesting physical effects and prospective optoelectronic applications. 
%
\end{abstract}

\pacs{Valid PACS appear here}
\keywords{Suggested keywords}
\maketitle


\section{Introduction}
One of the recent trends in contemporary nanotechnology is the use of van der Waals heterostructures for band structure engineering\cite{Geim,Novoselovaac9439}. This trend closely followed the seminal work on exfoliation of graphene with its unique electronic properties, which also provided a major boost to carbon-based optoelectronics. Exfoliation of graphene was in turn preceded by over a decade of extensive study of carbon nanotubes, following the pioneering work of Iijima\cite{Iijima} which still dwarfs any other publication on carbon-based nanostructures in the number of citations. In this paper we attempt to marry van der Waals heterostructures band-structure engineering with the rich physics of carbon-nanotubes in an attempt to design a self-organized new metamaterial with hyperbolic dispersion. 
 
Carbon nanotubes (CNTs) have been studied extensively since their discovery~\cite{Iijima}. The search for new applications based on their unique mechanical and electronic
properties is intensifying.  In particular, the high and gate-tunable electrical conductivity of CNT-based quantum wires provides a potential solution for on-chip interconnects and transistors in future integrated circuits~\cite{doi:10.1021/jz100889u}. Depending on the community-specific interests and targeted applications, nanotubes are considered as either single molecules or quasi-one-dimensional crystals with translational periodicity~\cite{RevModPhys.79.677}.

Electronic properties of single-walled CNTs are fully determined by the way they are rolled from a graphene sheet as described in detail in the vast  number of papers, reviews and textbooks. In what follows, we use the most common notations for CNT rolling adopted from Refs.~\cite{PhysRevB.51.11176, saito1998physical}

Recently, significant progress has been achieved in the controlled growth of horizontally aligned carbon nanotube (HACN) arrays~\cite{Nature_tech,C7CS00104E,Ichinose2019,Green2019,Gao2019}. Such structures show great potential as building blocks for transparent displays, nano electronics, quantum communication lines, field emission transistors, superstrong tethers, aeronautics and astronautics materials. CNT optical properties have been utilized in highly-efficient HACN-based terahertz (THz) polarizers~\cite{KonoNL09, KonoNL12} and optically-excited THz emitters~\cite{Hartmann,KonoNL15,Hartmann2019}. Other notable features of HACN arrays include giant mid-infrared inter-subband transitions~\cite{Kazuhiro} and recently predicted exciton ground state brightening~\cite{2017arXiv171002764} in the polarization-sensitive strong coupling regime~\cite{Kono18NP}.

Optical properties of planar HACN arrays have been a subject of extensive research; whereas, the in-plane electronic transport, which should be highly anisotropic, has not been studied. In particular, it would be interesting to look at the low-energy electronic spectrum of an array of metallic nanotubes. Indeed, graphene with its linear quasiparticle energy spectrum has become a popular ``playground'' to study ultra-relativistic phenomena on the table-top. It could be expected that aligning metallic single-walled carbon nanotubes in a regular array may produce an even more interesting dispersion. In this respect, zigzag $(3p,0)$ CNTs should be more promising than armchair tubes. Indeed, $(3p,0)$ nanotubes are in fact quasi-metallic with tiny curvature-induced gaps~\cite{PhysRevLett.78.1932}, which are highly sensitive to the environment. In addition, the energy spectrum degeneracy near the ``double'' Dirac point in zigzag CNTs, which stems from the graphene valley degeneracy and the way these nanotubes are rolled, promises a significantly richer in-plane dispersion in a regular array in comparison to the case of armchair nanotubes, in which the two Dirac points are well-separated in momentum space. 
For armchair CNTs, the main effect of introducing periodicity in the direction normal to the nanotube axis should be the opening of a small band gap, whereas degeneracy lifting in zigzag tubes is expected to result in a very complex dispersion stemming from crossings and avoided crossings of various energy level. Therefore, our study will be mostly focused on narrow-gap zigzag CNTs, with only limited results for armchair tubes of similar diameter presented mostly for comparison.         
Our approach to this problem is based on ab-initio calculations, which allows us to  simultaneously find the optimal distance between nanotubes in the array which minimizes the total energy and the in-plane energy spectrum. Ab-initio methods are widely used to study electronic properties of single carbon nanotubes with different chiralities~\cite{PhysRevB.65.155411,Antonov,UMENO200433} and various complex systems based on carbon nanotubes~\cite{RevModPhys.79.677,Osadchy}. On the basis of our ab-initio calculations we introduce a low-energy effective Hamiltonian, which describes the band structure of our system, and determines its parameters. The density of states is found for these parameters.

\begin{figure}[htb]
\centering
\includegraphics[scale=0.3,clip]{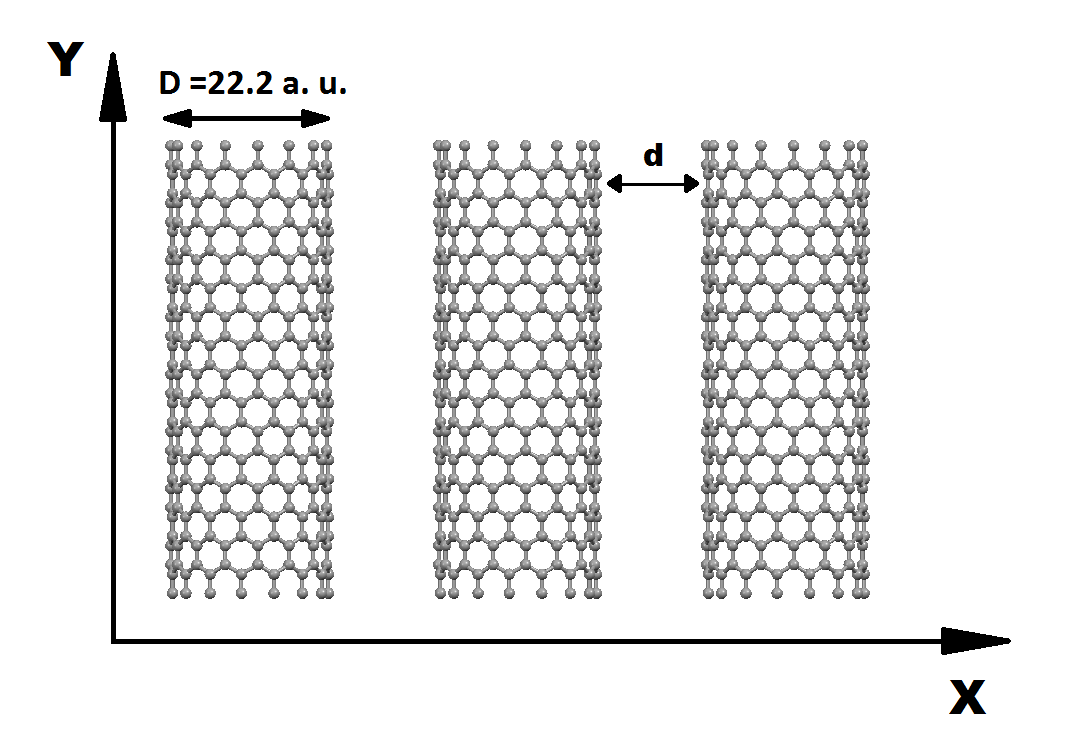}
\caption{
A horizontally aligned array of $(15,0)$ carbon nanotubes separated by a distance $d$. The diameter of a (15,0) CNT is 22.2 a.u. (1\,a.u=0.529\,\AA). 
\label{array}
}
\end{figure}

Our main aim is to calculate the electron dispersion of low-energy excitations in a perfect planar array of zigzag quasi-metallic carbon nanotubes. The tubes are assumed to be separated by the constant distance $d$ between the walls of the tubes (see Fig.~\ref{array}). In other words, the array is generated by creating a succession of exact copies of a carbon nanotube with separation along a direction normal to the tube axis. We have found the optimal distance between tubes using PBE-D2 approach (see Appendix~\ref{A1}). Our analysis includes calculation of the band structure of an isolated $(15,0)$ zigzag nanotube and an array of $(15,0)$ nanotubes separated by the optimal distance. 

\section{Band structure of an isolated $(15,0)$ carbon nanotube}

\begin{figure}[htb]
\centering
\includegraphics[scale=0.3,clip]{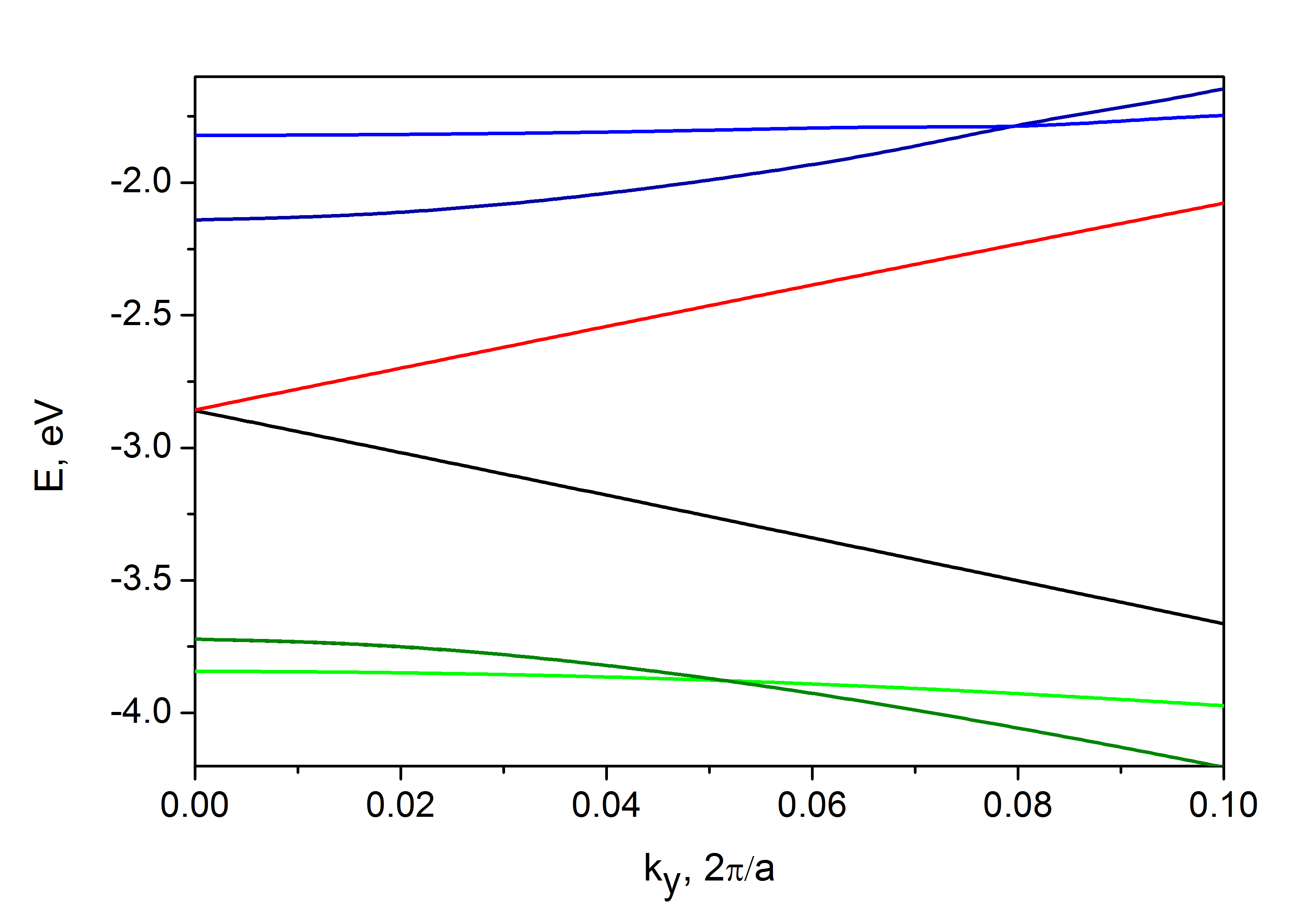}
\caption{The electron dispersion in a $(15,0)$ carbon nanotube along the tube axis for a small part of the Brilloin zone. The wave vector $k_{y}$ is given in the reciprocal space units. Here, $a$ is the translation vector, which is equal to $8$\,a.u. All the bands are two-fold degenerate. There is a small (few meV) curvature-induced band gap at $k_y =0$.}
\label{singleCNT}
\end{figure}

The low-energy part of the band structure of a single $(15,0)$ zigzag nanotube is shown in Fig.~\ref{singleCNT}. We have checked that our results are consistent  with the well-known ab-initio calculations of Ref.~\cite{doi:10.1021/jz100889u} which in turn are in good agreement with the experimental data of Lieber et.al.~\cite{Lieber} and confirm the $E_g \propto 1/D^3$ dependence for quasi-metallic CNTs with curvature-induced band gaps~\cite{PhysRevB.65.155411}.    

As expected, all the energy levels are two-fold degenerate (or four-fold degenerate when the Kramers degeneracy is taken into account). For single-tube calculations we used the Quantum Espresso code for a three-dimensional system with widely-separated nanotubes and confirmed that the dispersion in any direction normal to the nanotube axis was flat.

\section{Optimization of the nanotube array geometry}

Before calculating the optimal distance between nanotubes we performed the convergence test leading to the use of the ${10\times 10\times 1}$ Monkhorst-Pack grid (see Appendix~\ref{A1}). Then the PBE-D2 approach, which includes van der Waals interactions, was used to calculate the total energy of the  supercell for different values of the in-plane inter-tube separation $d$.

\begin{figure}[htb]
\centering
\includegraphics[scale=0.1,clip]{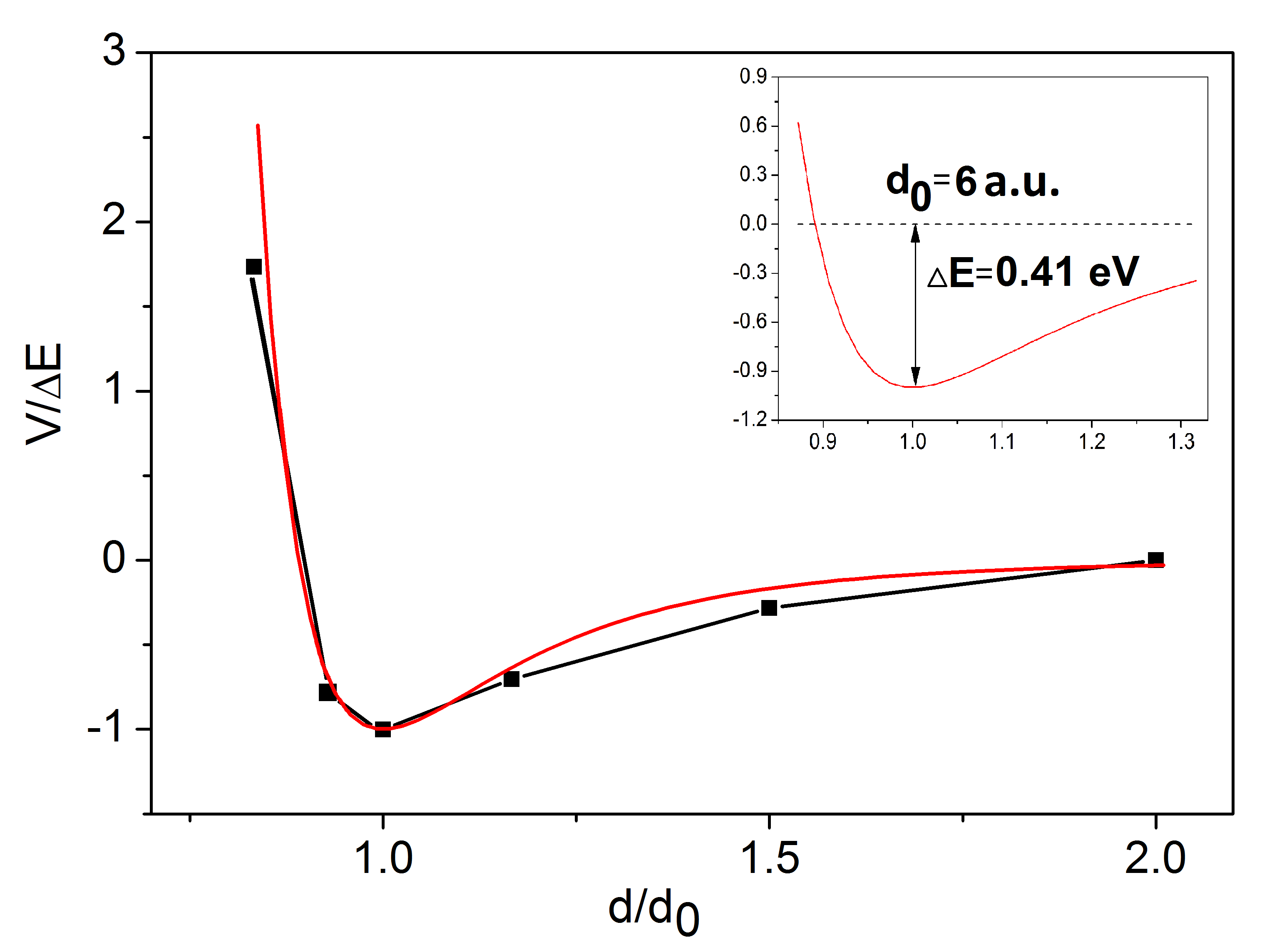}
\caption{The dependence of the  CNT array supercell total energy $V$ in units of the Lennard-Jones potential well depth $\Delta E$ on the normalized distance between the nanotube walls. A kinky dependence (shown in black) is the result of the PBE-D2 calculations; a smooth (red) curve represents the Lennard-Jones potential. The parameters of the best-fit Lennard-Jones potential are shown in the inset. Here $\Delta E$ is the depth of the potential well, and $d=d_0=6$\,a.u. corresponds to the potential minimum.} 
\label{compare}
\end{figure}

As can be seen from Fig.~\ref{compare}, the total energy dependence is well-described by the Lennard-Jones potential:
\begin{equation}
V(d) = \Delta E \left[-2\left(\frac{d_0}{d} \right)^6+
\left(\frac{d_0}{d}\right)^{12}\right],
\label{LJ}
\end{equation}
where $\Delta E$ is the depth of the Lennard-Jones potential well, and $d=d_0$ corresponds to the well minimum.

By fitting the results of our ab-initio simulations with the potential given by Eq.~(\ref{compare}), we find that the optimal inter-tube distance for the considered planar array of $(15,0)$ CNTs is given by $d=d_0=6$\,a.u.  

\section{Nanotube array energy spectrum}

\begin{figure*}[htb]
\centering
\includegraphics[scale=0.1,clip]{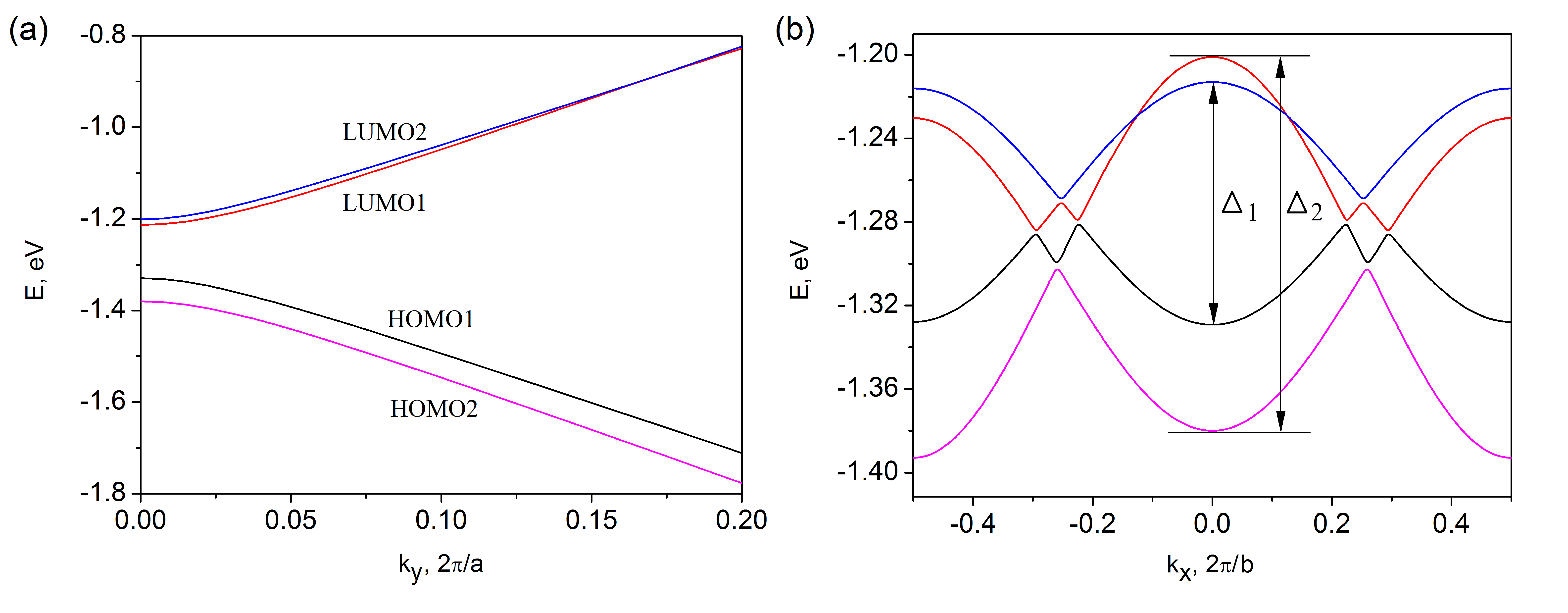}
\caption{The electronic band structure of an array of $(15,0)$ carbon nanotubes  separated by $d=6$\,a.u.: (a) along the tube axis for $k_x=0$; (b) in the direction normal to the tube axis for $k_y=0$. Here, $\Delta_1$ is the band gap between the LUMO1 and HOMO1 bands in the $\Gamma$ point; $\Delta_2$ corresponds to the LUMO2 and HOMO2 bands. The wave vectors $k_{y}$ and $k_{x}$ are given in reciprocal space units. The lattice period in $X$-direction $b=D+d=28.2$\,a.u. 
}
\label{bandarray}
\end{figure*}

We now proceed to the results of ab-initio calculations of the energy spectrum of the planar array of $(15,0)$ CNTs separated by $d=6$\,a.u. corresponding to the optimal inter-tube distance governed by the van der Waals interaction.
The calculated dispersions along the nanotube axis ($Y$-axis) and in the direction normal to the nanotube axis ($X$-axis) are shown in Fig.~\ref{bandarray}a and Fig.~\ref{bandarray}b, respectively.  

The striking feature of the calculated dispersion of the optimally-spaced CNT array is a significant gap (around $0.12$\,eV) opening in the $\Gamma$ point, in contrast to a tiny curvature-induced gap accompanied by a quasi-linear dispersion in the $Y$-direction and a totally flat dispersion in the $X$-direction in the limit of large inter-tube separation. The band gap shrinks towards the middle of the Brillouin zone in the $X$-direction. The dispersion along the $X$-axis resembles the well-known single chain tight-binding cosine spectrum with the period $b=D+d$, reflected in the horizontal axis due to electron-hole symmetry. 

An important feature of this spectrum is the lifting of the two-fold energy level degeneracy existing in a single zigzag CNT due to equivalent contributions from the two graphene $K$ points.
The most drastic consequence of lifting the valley-related degeneracy in quasi-metallic zigzag CNTs, when they are combined into a regular array, is the collapse of the band gap in the formed 2D material. As can be clearly seen from Fig.~\ref{bandarray}b, the two small gaps corresponding to the two formerly degenerate states of a single nanotube occur in the array at slightly different energies, which leads to an overlap of the conduction and valence bands. This unexpected dielectric-metal transition induced by the change in the system dimensionality seems to be a generic feature for closely-packed arrays of $(3p,0)$ CNTs. The results of similar calculations for the planar arrays of optimally-spaced $(12,0)$ and $(18,0)$ nanotubes are presented in Appendix~\ref{A2}. The band gaps are closed in these structures as well.         
Another notable feature of the zigzag nanotube array dispersion is the intersections of LUMO1 and LUMO2 bands which can be clearly seen in Fig.~\ref{bandarray}b near $k_x \approx \pm 0.125$ in units of $2\pi/b$. These intersections result in the appearance of two strongly tilted Dirac cones (or Dirac grooves), which annihilate when the separation between nanotubes tends to infinity and the degeneracy between the two bands is restored. In the literature, this type of energy bands intersection is usually called “type-II Dirac/Weyl point” or “tilted Dirac/Weyl cone with an open, hyperbolic isofrequency contour”, see e.g.~\cite{Mariani2018}. In 2D systems this type of dispersion leads to a plethora of interesting magnetotransport phenomena~\cite{PhysRevB.78.045415, C7CP03736H}. It has been discussed in relation to either functionalized graphene-like materials and strained graphene~\cite{StrainedGrapheneReview16} or artificial hexagonal lattices~\cite{Polini2013,Mariani2018} but has never been associated with carbon nanotube arrays. The discussed intersections are the consequence of the loss of the mirror symmetry, when the (3p,0) CNT with odd $p$ are placed in an array. 

\begin{figure}[h]
\centering
\includegraphics[scale=0.45,clip]{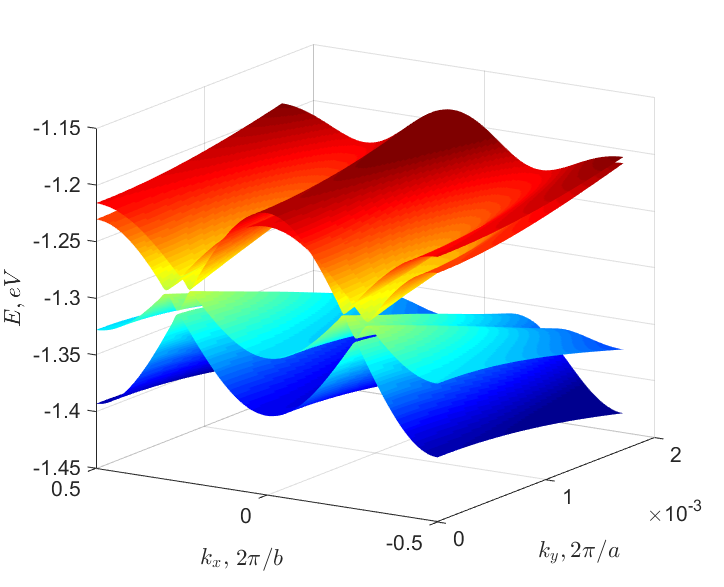}
\caption{
The surface plot of the energy dispersion of an array of $(15,0)$ carbon nanotubes  separated by $d=6$\,a.u. obtained within the PBE-D2 approach.}
\label{d6_surf}
\end{figure}

The two-dimensional surface plot of the four subbands closest to the Fermi level for the planar array of optimally-separated $(15,0)$ CNTS is shown in Fig.~\ref{d6_surf}. It clearly depicts the strong anisotropy of our system with different signs of quasiparticle effective masses in two orthogonal directions.

The main band structure parameters of the $(15,0)$ CNT array are summarized in Table~\ref{table1}. 

\begin{table*}
\begin{ruledtabular}
\caption{Band structure parameters of the optimally-spaced (15,0) CNT array. Indexes H1, H2, L1, L2 correspond to HOMO1, HOMO2, LUMO1, LUMO2 bands.}
\begin{tabular}{cccccc}
$V_{fx}, \frac{m}{s}$ & $m_{x}^{H2}, m_e$ & $m_{x}^{H1}, m_e$ & $m_{x}^{L1}, m_e $ & $m_{x}^{L2}, m_e$ & $\Delta_1, eV$ \\ \hline
$ 1.35\cdot10^5$             & 0.4         & 0.52      &-0.56    &-0.32 &0.12 \\ \hline \hline 
$V_{fy}, \frac{m}{s}$ & $m_{y}^{H2}, m_e$ & $m_{y}^{H1}, m_e$ & $m_{y}^{L1}, m_e $ & $m_{y}^{L2}, m_e$ & $\Delta_2, eV$ \\ \hline
$10.3\cdot10^5  $      &-0.068          & -0.071 & 0.07    &0.072   &0.18 \\ 
\end{tabular}
\end{ruledtabular}
\label{table1}
\end{table*}

\begin{figure*}[htb]
\centering
\includegraphics[scale=0.1,clip]{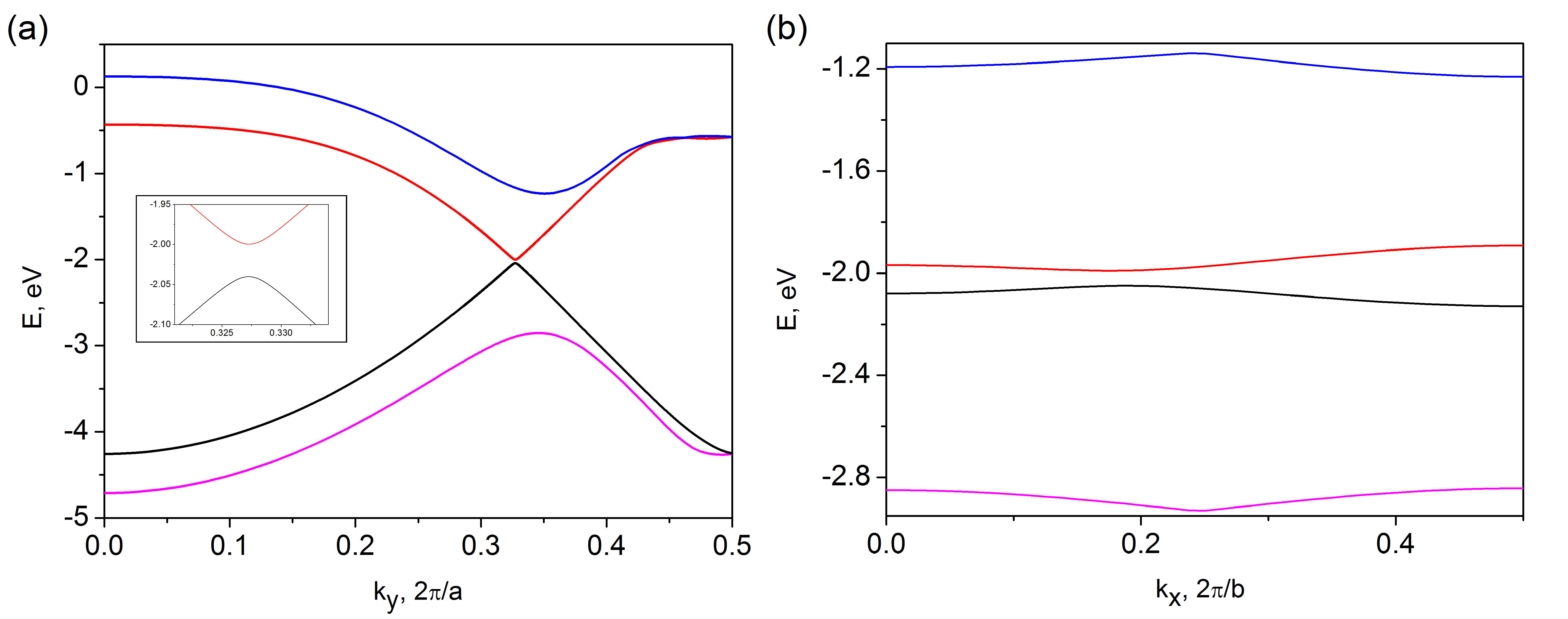}
\caption{The electronic band structure of an array of $(9,9)$ carbon nanotubes separated by $d=6$\,a.u.: (a) along the tube axis for $k_x=0.2$; (b) in the direction normal to the tube axis for $k_y=0.33$. The wave vectors $k_{x}$ and $k_{y}$ are given in reciprocal space units. The lattice period in the $X$-direction is $b=D+d=29.07$\,a.u. The inset to (a) shows the energy dispersion along the $Y$-axis in the close vicinity of the conduction band minimum.}
\label{armchair}
\end{figure*}

For comparison, we present in Fig.~\ref{armchair} the results of the ab-initio energy spectrum calculations for an optimally-spaced array of armchair $(9,9)$ nanotubes, which have a diameter, $D\approx 23.07$\,a.u., very close to that of the $(15,0)$ CNTs considered above. Notably, the optimal separation between the $(9,9)$ armchair CNTs, $d=6$\,a.u., is the same as between $(15,0)$ nanotubes in their optimally-spaced array.
The main consequence of combining metallic armchair nanotubes into a closely-packed regular planar array is the opening of a significant band gap, so that the resulting 2D crystal becomes a dielectric (narrow-gap semiconductor). In fact, this effect is equivalent to piercing an armchair nanotube by a strong magnetic field along the CNT axis~\cite{Portnoi}. Our calculations show that the band gap in an optimally-spaced $(9,9)$ tube is $E_g=40$\,meV which corresponds to an effective magnetic field of $65$\,T. This gap opening should be accompanied by strong interband dipole transitions in the narrow range of photon energies near the band gap~\cite{Portnoi,PortnoiSMS08,Hartmann2019}.
The dispersion remains strongly anisotropic with a very heavy effective mass in the $X$-direction near the conduction band minimum, and displays hyperbolic behavior near $k_y=2\pi/3$ and either $k_x=0$ or $k_x=\pi/b$. However, the overall low-energy band structure for an array of armchair CNTs is significantly less complex than in the case of narrow-gap zigzag nanotubes.

It should be noticed that the opening of a band gap in armchair nanotubes assembled into dense 3D bundles has been discussed in the literature~\cite{PhysRevB.65.155411,FromReich8, FromReich21}. However, as shown in Ref.~\cite{PhysRevB.65.155411}, this band gap opening depends on the system geometry and does not occur in highly-symmetric hexagonal packing of tubes, so it is worth checking what happens in the 2D configuration.

\section{The model Hamiltonian}

With the insight from our ab-initio calculations, we have constructed an approximate four-band model Hamiltonian for an array of optimally-spaced gapless zigzag carbon nanotubes. The energy spectrum of the array has four Dirac cones at $k_y=0$ and the values of $k_x$ given by $k_x = \pm k_1$ and 
$k_x = \pm k_2$. Around these points, the dispersion is anisotropic such that the ratio of the Fermi velocities along the $Y$ and $X$ directions is 
$v_{Fy}/v_{Fx} \equiv \xi \neq 1$. It is also evident that the energy spectrum is split at the gamma point (${\bf{k}} = 0$). 
To model these characteristics, we propose the following effective Hamiltonian:
\begin{equation}
H = 
\begin{pmatrix}
 H_{I} & V  \\ V & H_{II}
\label{eq:hamilt}
\end{pmatrix},
\end{equation}

where 
\begin{equation}
H_I = \alpha_{1}
\begin{pmatrix}
 \epsilon_1 & K_-^1 \\
 K_+^1 & \epsilon_1
\end{pmatrix}+\begin{pmatrix}
 \beta_1 k_{x}^2 & 0 \\
 0 & \beta_1 k_{x}^2
\end{pmatrix} ,
\nonumber
\end{equation}

\begin{equation}
H_{II} = \alpha_{2}
\begin{pmatrix}
 \epsilon_2 & K_-^2 \\
 K_+^2 & \epsilon_2 
\end{pmatrix} +\begin{pmatrix}
 \beta_2 k_{x}^2 & 0 \\
 0 & \beta_2 k_{x}^2
 \end{pmatrix} ,
\nonumber
\end{equation}

\begin{eqnarray}
K_\pm^{1,2} &=& \left[(k_x - k_{1,2}) -\lambda_{1,2}(k_x - k_{1,2})^3\pm \frac{i\cdot {k_y}}{\xi}\right] 
\nonumber \\
&\times& \left[(k_x + k_{1,2}) -\lambda_{1,2}(k_x + k_{1,2})^3\pm \frac{i\cdot {k_y}}{\xi}\right],
\nonumber
\end{eqnarray}
\begin{equation}
V = 
\begin{pmatrix}
 \upsilon & 0 \\
 0 & \upsilon
\end{pmatrix}.
\nonumber
\end{equation}
Here, parameters $k_1$ and $k_2$ specify the positions of Dirac cones. The corresponding Fermi velocities are defined by $\alpha_1$ and $\alpha_2$.
Parameter $\xi$ defines the ratio of the Fermi velocities in the $X$ and $Y$ directions. The band splitting is taken into account using  parameters $\epsilon_1$, $\epsilon_2$, and $\upsilon$. Parameters $\lambda_1$ and $\lambda_2$ in front of the cubic terms are responsible for the bending of bands at the edges of the Brillouin zone. The electron-hole asymmetry, which is clearly seen in our DFT calculations, is described by parameters $\beta_1$ and $\beta_2$. The Hamiltonian, Eq.~(\ref{eq:hamilt}), approximates the low-energy part of the energy spectrum with a good accuracy.

\begin{figure}[htb]
\centering
\includegraphics[scale=0.5,clip]{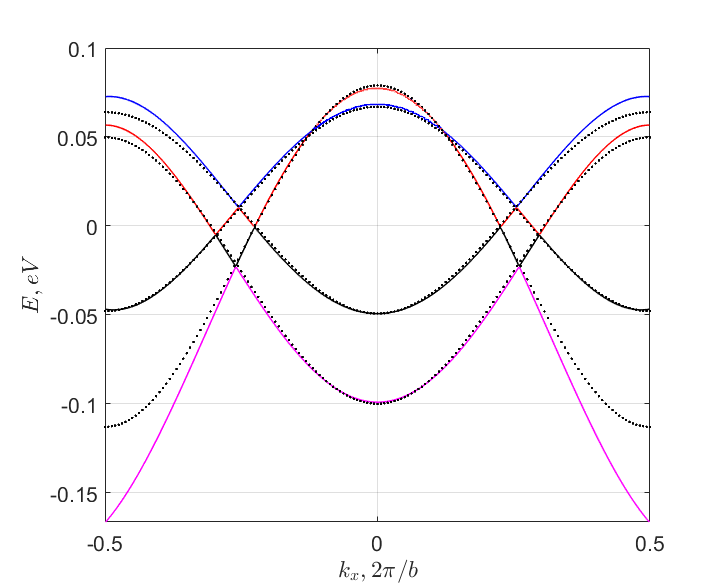}
\caption{The electronic band structure of the optimally-spaced ($d=6$\, a.u.) array of $(15,0)$ zigzag nanotubes in the direction normal to the tube axis. Solid lines correspond to band structure described by the effective Hamiltonian, dotted lines show the results of ab-initio calculations. Parameters of the model Hamiltonian are chosen in accordance with Table~\ref{table2}.}
\label{fig:2Dspectr}
\end{figure}

\begin{table*}
\centering
\caption{Parameters of the effective Hamiltonian fitted to ab-initio calculated electronic band structure for an array of optimally-spaced $(15,0)$ CNTs. The splitting constant $\upsilon$ is equal to $10^{-6}$\,a.u. The atomic system of units is used throughout the table: $\hbar = |e| = m_e = 1$.}
\begin{ruledtabular}
\begin{tabular}{cccccccccc}
 $\alpha_1$&$k_1, \frac{2\pi}{c}$&$\epsilon_1, a.u.$&$\lambda_1, {a.u.^{-1}}$
&$\beta_1$&$\alpha_2$&$k_2, \frac{2\pi}{c}$&$\epsilon_2, a.u.$&$\lambda_2, {a.u.^{-1}}$
&$\beta_2$\\ \hline
$0.039$&$0.255$&$0.00035$&$1.18$
&$0.0005$&$0.055$&$0.26$&$-0.0004$&$0.98$
&$-0.0065$\\ 
\end{tabular}
\label{table2}
\end{ruledtabular}
\end{table*}

The calculated energy bands from the ab-initio study of the CNT array described above are fitted by the dispersion given by the low-energy effective Hamiltonian~(\ref{eq:hamilt}). In Fig.~\ref{fig:2Dspectr}, we compare the results of the ab-initio calculations with the proposed few-parameter approximation. The relevant parameters for the effective Hamiltonian are given in Table~\ref{table2}. Polynomials of order higher than three should be taken into account in the Hamiltonian to get a more accurate dispersion.

\section{The density of states}

The band structure of the considered CNT array obtained with the help of the DFT approach has four saddle points. A simple hyperbolic energy spectrum around a saddle point with energy $E_0$ has the form
\begin{equation}
E(k_x,k_y) = E_0 + \frac{\hbar^2k_x^2}{2m_x} - \frac{\hbar^2k_y^2}{2m_y},
\label{eq:kspectrumgeneral}
\end{equation}
where $m_x$ and $m_y$ are the effective masses along the $X$ and $Y$ directions, respectively, 
which are defined as positive quantities. The effective masses are proportional to the inverse of the second derivative along the respective directions.

\begin{figure}[htb]
\centering
\includegraphics[scale=0.45,clip]{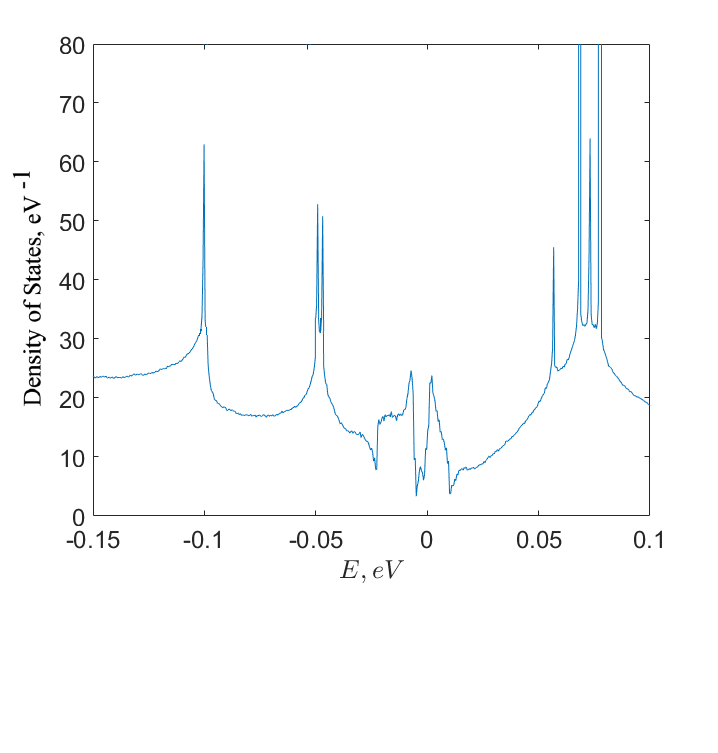}
\caption{The density of states of an array of optimally-spaced $(15,0)$ carbon nanotubes  calculated using the effective Hamiltonian given by Eq.~(\ref{eq:hamilt}). The sharp peaks correspond to Van Hove singularities.}
\label{dos}
\end{figure}

A calculation of the density of states (DOS) near a saddle point is outlined in Appendix \ref{A3}.
The result is that the DOS diverges in the limit $E\rightarrow E_0$ as $\frac{1}{2}\ln[1/(E-E_0)]$. These divergences give rise to peaks known as Van Hove singularities \cite{PhysRev.89.1189}, which manifest themselves in peaks in the optical absorption of the system. The density of states of an array of $(15,0)$ CNTs separated by $d=6$\,a.u., calculated from the spectrum given by the effective Hamiltonian~(\ref{eq:hamilt}) is shown in Fig.~\ref{dos}. The peaks in this figure correspond to saddle points of the energy spectrum, as can be seen by comparing Figs.~\ref{fig:2Dspectr} and \ref{dos}.

\section{Conclusions}

We have used ab-initio calculations to find the optimal (corresponding to the lowest total energy) geometry of a planar array of carbon nanotubes and to investigate the electron energy spectrum in the resulting super-structure. Detailed results are presented for an array of $(15,0)$ CNTs only; however, our further study shows that the main features of the spectrum, such as the closing of the total band gap, the appearance of tilted Dirac cones, strong anisotropy and the areas of negative dispersion near the Fermi level, are common for similar systems containing quasi-metallic zigzag nanotubes with a different diameter. These features are caused by lifting the valley-related degeneracy when zigzag CNTs are combined in a regular planar array. In contrast, combining metallic armchair nanotubes into an optimally-spaced planar array results in the opening of a significant band gap.

For an array of narrow-gap zigzag CNTs, we propose a semi-analytic $4 \times 4$ Hamiltonian describing the low-energy part of the energy spectrum with its distinctive hyperbolic dispersion.
The energy scale of the predicted hyperbolic dispersion makes the considered system a promising candidate for prospective applications in far-infrared optoelectronics. Indeed, the regions of negative effective mass in the low-energy part of the spectrum should lead to the appearance of negative differential conductivity, which can be used for generation of stimulated far-infrared radiation\cite{JETP_Lett99, JETP01}. The efficiency of infrared emission should be further enhanced by the presence of Van Hove singularities in the relevant spectral region which can be clearly seen from our DOS calculations. The considered structures can also be regarded as metasurfaces for electromagnetic waves in the optical to mid-infrared frequency range, which have been a subject of considerable recent research efforts\cite{Kildishev,Iorsh}.

The electronic properties of the considered system can be tuned by subjecting it to a strong magnetic field. Magnetic field behavior should reflect the system's extreme anisotropy. The component of the magnetic field normal to the array plane will result in non-trivial Landau quantization for hyperbolic materials. The influence of the in-plane component is envisaged to be mostly via the tuning of the nanotube band gap\cite{Smirnov, Portnoi}. Therefore, a strong dependence of the electron spectrum on the magnetic field direction should be expected.

We hope that our predictions of hyperbolic dispersion, tilted Dirac cones and geometry-induced metal-dielectric transitions in a planar array of optimally spaced carbon nanotubes will attract significant interest from material scientists to this hitherto overlooked fascinating 2D van der Waals material.  Clearly, it is a very promising system for both new exciting physics and future applications.

\section{Acknowledgement}
We acknowledge support from Horizon 2020 Marie Skłodowska-Curie RISE projects CoExAN (grant No 644076) and TERASSE (grant No 823878), Russian Mega-grant No 14.Y26.31.0015 and Project 3.2614.2017/4.6 of the Ministry of Education and Science of Russian Federation. This work was also financially supported by Government of Russian Federation (Grant 08-08) including the support for MEP through the ITMO Fellowship and Professorship Program. We are grateful to S.P.~Hepplestone, T.P.~Collier and E. Mariani for valuable comments.

\appendix

\section{Methods}\label{A1}

The ab-initio calculations of the band structure of the CNT array were performed with an ultra-soft pseudo-potential and a plane-wave basis in the Quantum Espresso (QE) package~\cite{giannozzi09}. The QE package is based on first principles density functional theory (DFT)~\cite{PhysRev.136.B864,PhysRev.140.A1133}. The DFT method is a powerful tool for performing many-body electronic structure calculations. It finds a good approximate solution to the quantum mechanical many-body problem. 
  
At the first stage, we considered a single $(15,0)$ zigzag carbon nanotube, which is metallic within the graphene zone-folding approximation. We examined the electronic bands of the nanotube along its axis. For the case of a $(15,0)$ CNT the unit cell contains 60 atoms. To employ the QE code, a simulated nanotube was placed in a hexagonal supercell with a lattice constant of 60\,a.u.  
A diameter of an $(m,n)$ single-walled carbon nanotube is computed in a.u. from the expression~\cite{saito1998physical}: $D=1.48\sqrt{ m^{2}+n^{2}+mn}$, where $n$ and $m$ are the integer components of the chiral vector. For a $(15,0)$ CNT, the diameter is equal to $22.2$\,a.u. The exchange-correlation energy functional of Perdew, Burke and Ernzerhof (PBE)~\cite{pbe96} was used. The ${1\times 4\times 1}$ Monkhorst-Pack grid in $k$-space was used for self-consistent calculations of electronic structure.

At the second stage, we made a convergence test with respect to the number of $k$-points in the ${n\times n\times 1 }$ Monkhorst-Pack grid for an infinite array of $(15,0)$ CNTs and different distances between tubes, varying $n$ from 4 to 12. We used the generalized gradient approximation~\cite{PhysRevB.46.6671} with PBE and added the van der Waals (vdW) corrections to find the optimal distance between tubes. The vdW interaction, which governs the optimal distance between CNTs, was included by the use of the method of Grimme (PBE-D2)~\cite{JCC:JCC20495}. In order to get accurate results, the plane-wave cutoff was set to a high value of $80$\,Ry, for which the structural relaxations and the electronic energies are fully converged. A Gaussian smearing for the occupations was used with a width of $0.01$\,eV. 
 
At the third stage, we calculated the band structure of the array of $(15,0)$ zigzag CNTs separated by the optimal distance $d$ found in the second stage. The nanotube array is assumed to lie in the $XY$-plane, and the tubes are aligned along the $Y$-axis, see Fig.~\ref{array}, so that the system is two-dimensional in nature. To adapt our system to the standard QE package we assume the CNTs to be arranged into the three-dimensional lattice with the large separation of 60~a.u. between the layers in the $Z$-direction, so that the interaction between the tubes in this direction can be neglected. The band structure calculations were performed within the above mentioned DFT (PBE-D2) method. 

\section{Energy spectra for optimally-spaced arrays of (12,0) and (18,0) carbon nanotubes}\label{A2}

In this section we provide some results of our ab-initio calculations supporting the statement in the main text of the paper that the collapse of the band gap is a generic feature of closely-packed arrays of $(3p,0)$ zigzag carbon nanotubes.

Whereas the main text is focused on $(15,0)$ CNTs, here we present the data for two other quasi-metallic zigzag CNTs, which have either a smaller or a larger diameter than the $(15,0)$ CNT. Namely, we present here the data for optimally-spaced arrays of $(12,0)$ and $(18,0)$ CNTs. The calculations are performed within the PBE-D2 approach described in the previous section.

\begin{figure}[htp]
\centering
\includegraphics[scale=0.35,clip]{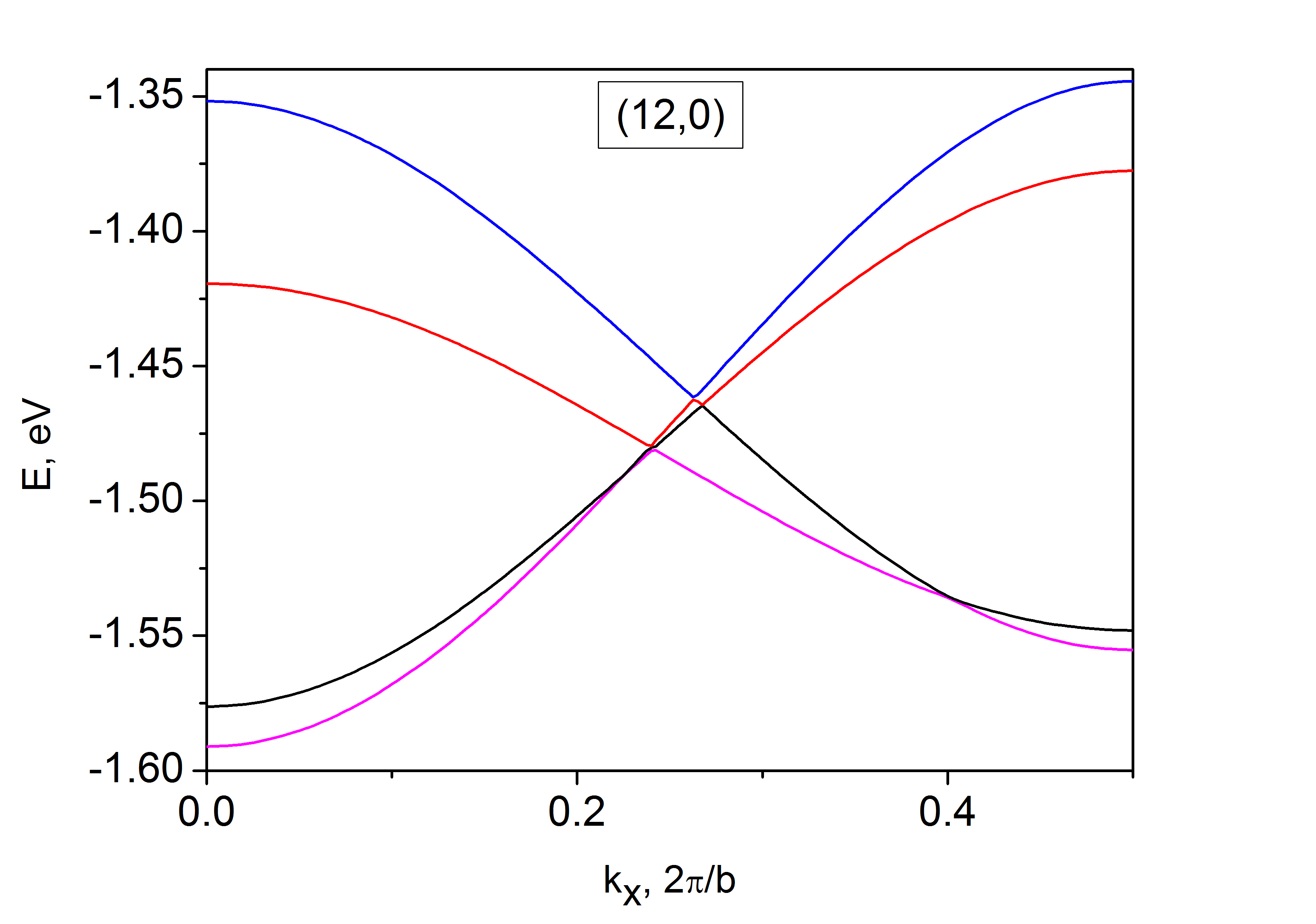}
\caption{The electronic dispersion of an array of $(12,0)$ carbon nanotubes separated by $d=6.2$\,a.u. in the direction normal to the tube axis for $k_y=0$. The lattice period in the $X$-direction $b=D+d=23.96$\,a.u.}  
\label{CNT12_0}
\end{figure}

In Fig.~\ref{CNT12_0}, we present the electronic dispersion $E(k_x)$ in the planar array of $(12,0)$ CNTs along the direction normal to the nanotube axis for $k_y=0$.

\begin{figure}[htp]
\centering
\includegraphics[scale=0.35,clip]{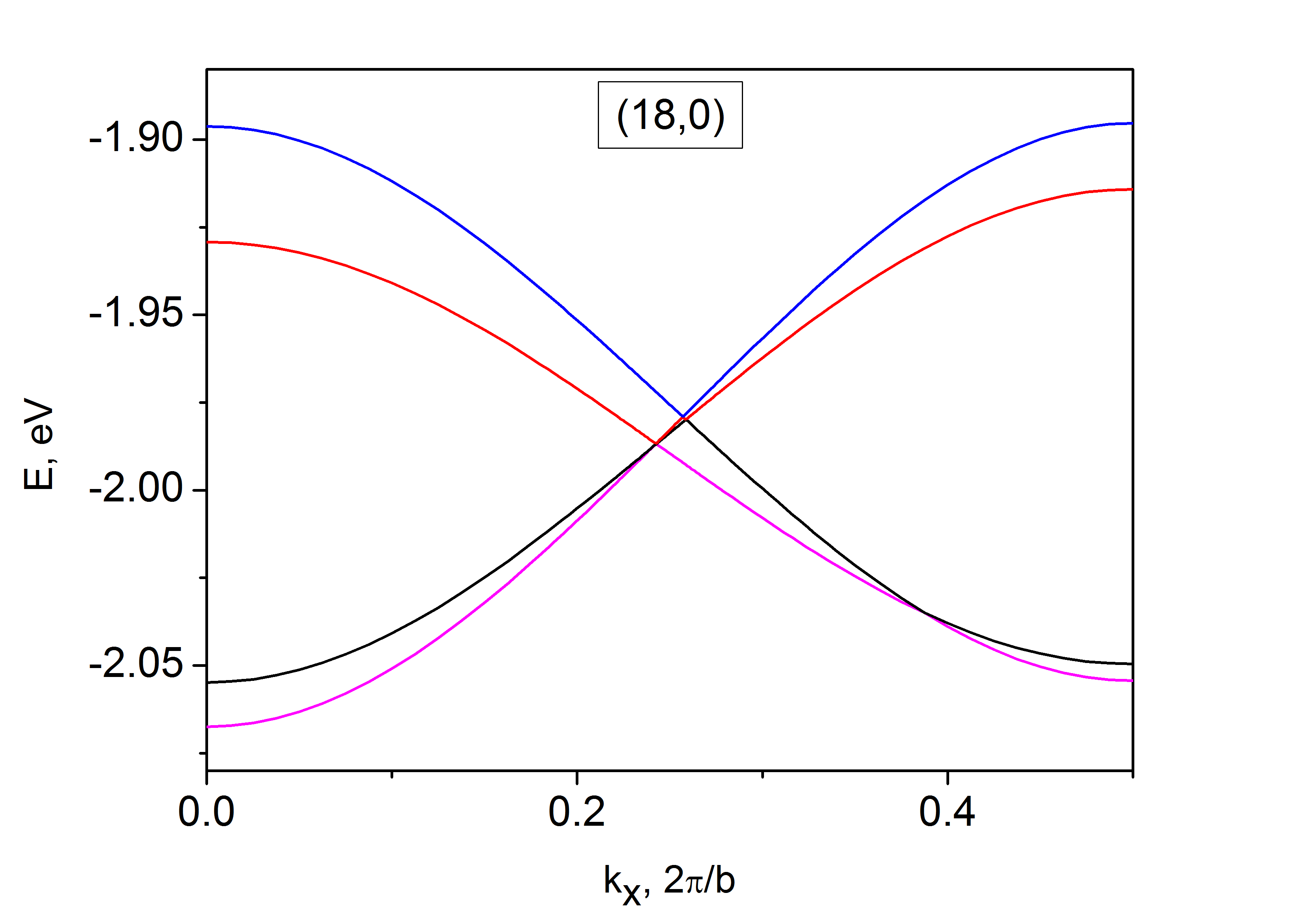}
\caption{The electronic dispersion of an array of $(18,0)$ carbon nanotubes separated by $d=6.2$\,a.u.  in the direction normal to the tube axis for $k_y=0$. The lattice period in $X$-direction $b=D+d=32.84$\,a.u.} 
\label{CNT18_0}
\end{figure}

A similar plot for optimally spaced planar array of $(18,0)$ CNTs is shown in Fig.~\ref{CNT18_0}. Interestingly, the optimal nanotube separation $d$ in this array is the same as for $(12,0)$ CNTs which exceeds slightly the value for $(15,0)$ nanotubes.   

Clearly, in both cases there is no value of energy in the low-energy part of the spectrum for which the valence band and conduction band of the zigzag-nanotube-based 2D crystal do not overlap. This indicates a metallic behavior of an array made of nanotubes which individually have curvature-induced band gaps of a few meV.   

\section{Density of states in a hyperbolic material}\label{A3}

Here, we calculate the density of states, $g(E)$, near a general saddle point, where the energy zero level is chosen to coincide with the saddle point $E_0$. The energy is then given by
\begin{equation}
E(k_x,k_y) = \frac{\hbar^2k_x^2}{2m_x} - \frac{\hbar^2k_y^2}{2m_y}.
\label{eq:kspectrum}
\end{equation}
where the positive constants $m_x$ and $m_y$ are the effective masses along the $X$ and $Y$ directions, respectively.

By definition, the density of states is given by 
\begin{equation}
g(E) = \frac{2}{(2\pi)^2}\int_{E(\mathbf k) = \textrm{const.}}{\frac{df_E}{|\nabla _{\mathbf k}E(\mathbf k)|}}
\end{equation}
where $df_E$ is a line element along the constant energy curve~\cite{Ibach2009, Ashcroft}, which we now proceed to find. Dividing both sides of Eq.~(\ref{eq:kspectrum}) by $E$, we arrive at 
\begin{equation}
\frac{k_x^2}{a^2} - \frac{k_y^2}{b^2} = 1
\end{equation}
which defines a hyperbola with

\begin{equation}
a = \frac{\sqrt{2m_xE}}{\hbar}, 
\nonumber \\
b = \frac{\sqrt{2m_yE}}{\hbar}.
\nonumber
\end{equation}
The constant energy curves are thus two hyperbolas. The parametric representation of a wavevector $\mathbf k = [k_x, k_y]^T$ along a hyperbola is given by, 
\begin{equation}
k_x(t) = a\cosh t, \nonumber\\
k_y(t) = b\sinh t,
\end{equation}
where $t \in [-t_0,t_0]$, with the cut-off parameter $t_0$ defined by the condition

\begin{equation}
t_0
 =\min\left\{\ln\left(\frac{k_{0x}}{a} - \sqrt{\frac{k_{0x}^2}{a^2}-1}\right), 
  ~\ln\left(\frac{k_{0y}}{b} - \sqrt{\frac{k_{0y}^2}{b^2}+1}\right)
 \right\}
\label{eq:t0}
\end{equation}
Now, using the standard technique for evaluating a line integral, we calculate
\begin{widetext}
\begin{eqnarray}
g(E) &=& \frac{1}{2\pi^2}\int_{-t_0}^{t_0}\frac{|\dot{\mathbf k}(t)|}{|\nabla_\mathbf{k}E(\mathbf k(t))|}dt 
 = \frac{1}{2\pi^2}\int_{-t_0}^{t_0}\frac{b\sqrt{\cosh^2t+\eta\sinh^2 t}}{\frac{\hbar^2\sqrt{\eta}b}{m_x}\sqrt{\cosh^2 t + \eta \sinh^2 t}}dt 
 \nonumber \\
&=& \frac{1}{2\pi^2}\int_{-t_0}^{t_0}\frac{m_x}{\hbar^2\sqrt{\eta}}dt
= \frac{t_0\sqrt{m_xm_y}}{(\pi\hbar)^2},
\end{eqnarray}
\end{widetext}
where 
\begin{equation}
\eta \equiv \frac{m_x}{m_y} = \frac{a^2}{b^2}. 
\end{equation}
For small $E$, Eq.~(\ref{eq:t0}) yields
\begin{equation}
t_0 \approx \min\left\{\ln\left(\frac{2\hbar k_x^0}{\sqrt{2m_xE}}\right), ~\ln\left(\frac{2\hbar k_y^0}{\sqrt{2m_yE}}\right) \right\}.
\end{equation}
This means that $t_0$ and therefore $g(E)$ diverges in the limit $E\rightarrow 0$ as $\frac{1}{2}\ln(1/E)$.


\bibliography{bibl}


\end{document}